\documentclass[12pt]{llncs}

\usepackage{amsmath}
\usepackage{amssymb}
\usepackage{epsfig}
\usepackage{afterpage}
\usepackage{fullpage}

\def\Proof{\par\noindent{\bf Proof:}\indent}
\def\QED{\hfill$\Box$\par\vskip1em}
\def\PROG#1{$\mathcal{#1}$}

\begin{document}

\title{Infinite Unlimited Churn}

\author{Dianne Foreback\inst{1} \and 
		 Mikhail Nesterenko\inst{1} \and
         S\'{e}bastien Tixeuil\inst{2}}

\institute{Kent State University, Kent, OH, USA \and UPMC Sorbonne Universit\'{e}s \& IUF, Paris, France}

\date{}
\maketitle
\thispagestyle{plain}

\begin{abstract}
We study unlimited infinite churn in peer-to-peer overlay
networks. Under this churn, arbitrary many peers may concurrently
request to join or leave the overlay network; moreover these requests
may never stop coming. We prove that unlimited adversarial churn,
where processes may just exit the overlay network, is unsolvable. We
focus on cooperative churn where exiting processes participate in the
churn handling algorithm. We define the problem of unlimited infinite
churn in this setting.  We distinguish the fair version of the
problem, where each request is eventually satisfied, from the unfair
version that just guarantees progress. We focus on local solutions to
the problem, and prove that a local solution to the Fair Infinite
Unlimited Churn is impossible. We then present and prove correct an
algorithm \PROG{UIUC} that solves the Unfair Infinite Unlimited Churn
Problem for a linearized peer-to-peer overlay network. We extend this
solution to skip lists and skip graphs.
\end{abstract}

\section{Introduction}

In a peer-to-peer overlay network, each member maintains the
identifiers of its overlay neighbors in its memory while leaving
message routing to the underlay. A peer-to-peer overlay network is
inherently decentralized and scales up easily. Peer-to-peer overlays
are well suited for distributed content storage and delivery. Recent
applications of such overlays to internet telephony~\cite{skype} and
digital cryptocurrencies~\cite{bitcoin} further enhance interest in
scientific studies of the principles of peer-to-peer overlay networks.

Due to the lack of central authority and the volunteer nature of
overlay network participation, \emph{churn}, or joining and leaving of
peers, is a particularly vexing problem affecting peer-to-peer overlay
networks. Churn may be cooperative, if departing processes execute a
prescribed departure algorithm; or adversarial, if they just exit.

\ \\ \textbf{Infinite and unlimited churn.}  Every peer-to-peer
overlay network has to handle churn. Usually, while the topological
changes in the overlay required by the churn requests occur, the
primary services of the overlay (\emph{e.g.} content retrieval) are
either considered suspended or they are disregarded altogether.  In
other words, the churn is considered finite, and the overlay network
users just wait until join/leave requests stop coming, then the
overlay network recovers and restores services. If churn happens
again, the service suspension repeats. This may be tolerable if churn
is infrequent since the overlay network is available most of the time.
However, at the scales that peer-to-peer overlay networks achieve,
peers nearly always wish to join or leave the overlay network. In this
case, the service degradation caused by intermittent suspensions may
become unacceptable. That is, it is necessary to consider
\emph{infinite churn} under which the overlay network has maintain
services while handling it.

One way to handle churn is to engineer sufficient redundancy in the
overlay network topology, so that if the peers leave, there are enough
alternative paths to enable service operation. In this approach, the
amount of redundancy necessarily places a limit on the number of
processes that churn concurrently: the churning processes must not
sever all redundant paths. However, under heavy churn load and large
scale, a peer-to-peer overlay network may breach this limit and
collapse (that is, partition itself).  Also, the necessity to add
redundancy leads to wasted resources. Hence, there is an interest in
studying the possibility of \emph{unlimited churn}, where this is no
bound on the number of concurrently joining or leaving processes.

This paper is an attempt to define and study infinite unlimited churn
in peer-to-peer overlay networks.

\ \\ \textbf{Unfair and local churn.} A request to join and, in
cooperative churn, leave the overlay network is submitted to the
overlay by the churning process. A churn handling algorithm is
\emph{fair} if it eventually satisfies every request. By contrast, a
churn algorithm that allows the possibility, under infinite churn, to
bypass indefinitely some requests (still guaranteeing progress, that
is, satisfying some churn requests indefinitely), is
\emph{unfair}. Unfair algorithms are possibly more efficient.

Potentially, a churn handling solution may be straightforward and
\emph{global}: have some distinguished process manage all churn
requests and process them. However, such a centralized global solution
is not practical for large scale overlay networks as it creates a
performance bottleneck and a single point of failure. In contrast, in
a \emph{local} solution, only processes in the immediate vicinity of
the churning process are involved in processing the request.

In this paper, we study fairness and locality of churn solutions.

\ \\ \textbf{Our contribution.}  Specifically, we consider churn in
the asynchronous message passing system model. We prove that there
does not exist an algorithm that can handle unlimited adversarial
churn. We then focus on cooperative unlimited churn. We define the
Infinite Unlimited Churn Problem by specifying the properties of a
churn handing algorithm. We distinguish fair and unfair types of the
problem. We prove that there is no local solution to the Fair Infinite
Unlimited Churn Problem. We then present an algorithm that solves the
unfair version of the problem while maintaining a linear topology,
i.e. topological sort. We extend our algorithm to handle skip lists
and skip graphs.

To the best of our knowledge, this paper is the first systematic study
of unlimited infinite churn.

\ \\ \textbf{Related work.}  Independently of peer-to-peer overlay
networks, several papers~\cite{ko2008using,mega2012churn,churn}
address determination of the rate of churn, which is a difficult task
itself. Fundamental problems in Distributed Computing, such as
Agreement, were also studied in the context of
churn~\cite{augustine2013fast,augustine2015enabling,guerraoui2013highly},
however their inherently global setting makes them unsuitable for
building peer-to-peer overlays.

Usually, peer-to-peer overlay networks are designed to have redundant
links so that they can withstand limited peer departure (that is,
limited adversarial churn in our
terminology)~\cite{hyperring,AwerbuchS07,Awerbuch2009,gambs2012scalable}.
Many papers address repairing the topology after determining a process
unexpectedly left the overlay
network~\cite{ron,augustine2015enabling,drees2016churn,hayes2012forgiving,pastry,saia2008picking}.
Limited churn also enables the possibility to maintain overlay
services while adjusting~\cite{augustine2013storage,kuhn2010towards}.

Another approach deals with self-stabilizing peer-to-peer overlay
maintenance
algorithms~\cite{Berns2011,CaronDPTjournal,Dolev2004,foreback2014stabilizing,linearizationLATIN,JRSST09,jacob2014skip+,koutsopoulosSS15universal,corona,linoracles,shaker2005self}
that enable the peer-to-peer overlay network to recover from an
arbitrary topology disruption, once it stops.  That is,
self-stabilizing algorithms may handle unlimited but finite
adversarial churn.  In these algorithms, churn is handled implicitly:
topology changes are assumed to eventually stop, after this the system
is designed to recover.  To handle the possibility of initial
incorrect state, the use of oracles (that is, abstract entities that
provide information about the overlay network) may be
necessary~\cite{foreback2014stabilizing,linoracles}. The recent trend
is to provide a general framework that can be instantiated for various
overlay networks~\cite{Berns2011,koutsopoulosSS15universal}.

Overall, previous studies in the context of peer-to-peer overlay
networks consider \emph{limited} and/or \emph{finite} churn, while
this paper focuses on unlimited and infinite churn.

\section{Model and Problem Statement}

\noindent
\textbf{Peer-to-peer overlay networks, topology.} A peer-to-peer
overlay network consists of a set of processes with unique
identifiers. We refer to processes and their identifiers
interchangeably.  Processes communicate by message passing.  A process
stores identifiers of other processes in its memory. Process $a$ is a
\emph{neighbor} of process $b$ if $b$ stores the identifier of
$a$. Note that $b$ is not necessarily a neighbor of $a$.  A process
may send a message to any of its neighbors.  Message routing from the
sender to the receiver is carried out by the underlying network.  A
process may send a message only to the receiver with a specific id,
i.e. we do not consider broadcasts or multicasts.  Communication
channels are FIFO with unlimited message capacity.

A \emph{structured} peer-to-peer overlay network maintains a
particular topology. One of the basic topologies is \emph{linear}, or
a topological sort, where each process $b$ has two neighbors $a < b$
and $b > c$ such that $a$ is the highest id in the overlay network
that less than $b$ and $c$ is the lowest id greater than $b$.
Consider a particular topology. A \emph{cut-set} is a (proper) subset
of processes of the network such that the removal of these processes
and their incident edges disconnects the network. It is known that if
a network topology is not a complete graph, it has a cut-set. Since a
peer-to-peer overlay network maintains its connectivity by storing
identifiers in the memory of other processes, once disconnected it may
not re-connect. Hence, a peer-to-peer overlay network must not become
disconnected either through the actions of the algorithm or through
churn actions.

\ \\ \textbf{Searching, joining and leaving the overlay network.}  The
primary use of a peer-to-peer overlay network is to determine whether
a certain identifier is present in the overlay network.  A search
request message bearing an identifier, may appear in the incoming
channel of any process that has already joined the overlay
network. The request is routed until either the identifier is found or
its absence is determined.

A process may request to join the overlay network. We abstract
bootstrapping by assuming that a join request, bearing the joining
process identifier, appears in an incoming channel of any process that
has already joined the overlay network.  A process that joined the
overlay network may leave it in two ways. In \emph{adversarial churn}
a leaving process just exits the overlay network without participating
in further algorithm actions. In \emph{cooperative churn} a leaving
process sends a request to leave the overlay network. The leaving
process exits only after it is allowed to do so by the algorithm.  A
process may join the overlay network and then leave. However, a
process that left the overlay network may not join it again with the
same identifier. A join or leave request is a \emph{churn request} and
the corresponding join or leave message is a \emph{churn message}.
When a leaving process exits the overlay network, the messages in its
incoming channels are lost.  However, the messages sent from this
process before exiting remain in the incoming channel of the receiving
process.

\ \\ \textbf{Infinite and unlimited churn definitions.} Churn is
\emph{infinite} if the number of churn requests in a computation may
not be bounded by a constant either known or unknown to the
algorithm. Churn is \emph{unlimited} if the number of concurrent churn
requests in the overlay network is not bounded by a constant either
known or unknown to the algorithm. Note that unlimited churn means
that potentially all processes that are presently in the overlay
network may request to leave. Note also that the two properties are
orthogonal. For example, churn may be finite but unlimited: all
processes may request to leave but no more join or leave requests are
forthcoming. Alternatively, in infinite limited churn, there may be a
infinite total number of join or leave requests but only, for example,
five of them in any given state.

In this paper we only consider infinite unlimited churn.

\ \\ \textbf{Churn algorithm.}  A churn algorithm handles churn
requests. For each process, an algorithm specifies a set of variables
and actions.  An \emph{action} is of the form $ \langle label\rangle:
\langle guard \rangle \longrightarrow \langle command \rangle$ where
\emph{label} differentiates actions, \emph{guard} is a predicate over
local variables, and \emph{command} is a sequence of statements that
are executed \emph{atomically}. The execution of an action transitions
the overlay network from one state to another.  An algorithm
\emph{computation} is an infinite fair sequence of such states. We
assume two kinds of fairness of computation: weak fairness of action
execution and fair message receipt. \emph{Weak fairness} of action
execution means that if an action is enabled in all but finitely many
states of the computation then this action is executed infinitely
often. \emph{Fair message receipt} means that if the computation
contains a state where there is a message in a channel, this
computation also contains a later state where this message is not
present in the channel, i.e. the message is received.  We place no
bounds on message propagation delay or relative process execution
speeds, i.e. we consider fully asynchronous computations.

\ \\ \textbf{Algorithm locality.} A churn request may potentially be
far, i.e. a large number of hops, from the place where the topology
maintenance operation needs to occur.  \emph{Place of join} for a
join request of process $x$, is the pair of processes $y$ and $z$ that
already joined the overlay network, such that $y$ has the greatest
identifier less than $x$ and $z$ has the smallest identifier greater
than $x$.  In every particular state of the overlay network, for any
join request, there is a unique place of join. Note that as the
algorithm progresses and other processes join or leave the overlay
network, the place of join may change. \emph{Place of leave} for a
leave request of process $x$ is defined similarly. \emph{Place of
  churn} is a place of join or leave.

A network topology is \emph{expansive} if there exists a constant $m$
independent of the network size such that for every pair of processes
$x$ and $y$ where the distance between $x$ and $y$ is greater than
$m$, a finite number of processes may be added $m$ hops away from $x$
to increase the distance between $x$ and $y$ by at least one. This
constant $m$ is the \emph{expansion vicinity} of the topology.  In other
words, in an expansive topology, every pair of processes far enough
away may be further separated by adding more processes to the network
without modifying the expansion vicinity of one member of the pair.
Note that a completely connected topology is not expansive since the
distance between any pair of processes is always one. However, a lot
of practical peer-to-peer overlay network topologies are expansive. For
example, a linear topology is expansive with expansion vicinity of
$1$ since the distance between any pair of processes at least two hops
away may be increased by one if a process is added outside the
neighborhood of one member of the pair.

A churn algorithm is \emph{local} if there exists a constant $l$
independent of the overlay network size, such that only processes within $l$
hops from the place of churn need to take steps to satisfy this churn
request. The minimum such constant $l$ is the \emph{locality} of the
algorithm. Note that a local algorithm may maintain only an
expansive topology, and that the expansive vicinity of this topology must be
greater than the locality of the algorithm.

\ \\ \textbf{The Infinite Churn Problem.} A \emph{link} is the state
of channels between a pair of neighbor processes.  As a churn
algorithm services requests, it may temporarily violate the overlay
network topology that is being maintained. A \emph{transitional link}
violates the overlay network topology while a \emph{stable link}
conforms to it.  An algorithm that solves the infinite churn problem
conforms to a combination of the following properties:

\begin{description}
\item[\em request progress:] if there is a churn request in
the overlay network, some churn request is eventually satisfied; 
\item[\em fair request:] if there is a churn request in the overlay network, this churn
request is eventually satisfied; 
\item[\em terminating transition:]
every transitional link eventually becomes stable;
\item[\em message progress:] a message in a stable link is either delivered or
forwarded closer to the destination; 
\item[\em message safety:] a message in a transitional link is not lost.
\end{description}

Note that the fair request property implies the request progress
property. The converse is not necessarily true.

%
%

\begin{definition}\label{defUnfair}
\emph{The Unfair Infinite Unlimited Churn Problem} is the combination
of request progress, terminating transition, message progress and
message safety properties.
\end{definition}

\begin{definition}\label{defFair}
\emph{The Fair Infinite Unlimited Churn Problem} is the combination
of fair request, terminating transition, message progress and message
safety properties.
\end{definition}

In other words, Fair Infinite Unlimited Churn guarantees that every churn
request is eventually satisfied while Unfair Infinite Unlimited Churn does not.

\section{Impossibilities}

\noindent\textbf{Adversarial churn.}

\begin{theorem}\label{trmNoAdversarial}
There does not exist a solution for unlimited adversarial churn if the
maintained topology is not fully connected.
\end{theorem}

\Proof Assume there is an algorithm \PROG{A} that maintains a topology
that is not fully connected such that \PROG{A} is resilient against
unlimited adversarial churn. Let $N$ be the number of processes in the
overlay network in some global state. If the topology is not fully
connected, there exists a cut-set $P$ of processes whose cardinality
is less than $N$. Since \PROG{A} handles unlimited churn, it should
handle the departure of every process in $P$. However, since $P$ is a
cut-set, its departure disconnects the overlay network, which is not
possible to handle. Hence, \PROG{A} is not able to handle unlimited
adversarial churn.  \QED

Since adversarial unlimited churn cannot be handled, in the rest of the paper
we are considering cooperative (unlimited) churn.

\ \\ \textbf{Fair local churn.}  Intuitively, the proof of the below
theorem describes the continuous servicing of join requests before a
distant churn request can reach its place of churn. This join chain
keeps building in front of the distant churn request, precluding it
from ever reaching the appropriate place.

\begin{theorem}\label{impossible}
There does not exist a local solution to the Fair Infinite Unlimited
Churn Problem for an expansive overlay network topology.
\end{theorem}

\Proof Assume there exists a local algorithm \PROG{A} with locality
$l$ that satisfies the Fair Infinite Unlimited Churn Problem while
maintaining an expansive topology with expansion vicinity of $m>l$. Let
us add a churn request with id $x$ to the incoming channel of another
process $y$ whose distance $d$ to the place of churn for $x$ is
greater than $m$ and, therefore, $l$. Since the topology is expansive,
there is a set of processes $Z$ that are not part of the overlay
network such that every process in this set can be added $m$ hops away
from $y$ such that the distance $d$ between $y$ and the place of churn
for $x$ increases by at least one. Observe that \PROG{A} is assumed
local with locality $l$ that is less than expansion vicinity $m$. This
means that every process in $Z$ may be added to the overlay network
without $y$ having to take a step. Let us add the processes of $Z$ to
the overlay network.  Let us then have $y$ receive the churn request
for $x$. Since \PROG{A} is a solution to the Fair Infinite Unlimited Churn
Problem, $y$ must forward this request closer to its place of
churn. Suppose $y$ forwards this request to some process $u$. However,
since the distance from $y$ to $x$'s place of churn is at least $d+1$,
the distance from $u$ to this place of churn is at least $d$. We
continue this procedure ad infinitum.

The resultant sequence is a computation of algorithm \PROG{A}. Yet,
there is a churn request of process $x$ that is never satisfied. This
means that \PROG{A} violates the fair request property of the Fair
Infinite Unlimited Churn Problem, which contradicts our initial assumption.
\QED

Since fair local (infinite unlimited) churn is impossible, in the next
section we address unfair local (infinite unlimited) churn.

\section{Unfair Local Infinite Unlimited Churn for a Linear Topology}

\noindent
\textbf{Linear topology under churn.}  In a linear topology, each
process $p$ maintains two identifiers: \emph{left}, where it stores
the largest identifier greater than $p$ and \emph{right}, where it
stores the smallest identifier less than $p$.  Processes are thus
joined in a chain. For ease of exposition, we consider the chain laid
out horizontally with higher-id processes to the right and lower-id
processes to the left.  The largest process stores positive infinity
in its \emph{right} variable; the smallest process stores negative
infinity in \emph{left}. A \emph{left end} of a link is the smaller-id
neighbor process. A \emph{right end} is the greater-id process.

As a process joins or leaves the overlay network it may change the
values of its own or its neighbors variables thus transitioning the
link from one state to another.  In a linear topology, a link is
\emph{transitional} if its left end is not a neighbor of its right end
or vice versa.  The link is \emph{stable} otherwise. The largest and
smallest processes may not leave. The links to the right of the
largest process and to the left of the smallest processes are always
stable. A process may leave the overlay network only after it has
joined. We assume that in the initial state of the overlay network,
all links are stable.

\afterpage{
\thispagestyle{empty}
\begin{figure}
\scriptsize
\begin{tabbing}
1\=123456789023\=12345\=12345\=12345\=12345\=12345\=12345\=12345\=12345\=12345\=\kill

\>\textbf{constant}\ $p$ //  process identifier \\
\>\textbf{variables} \\
\>\>$\mathit{left}, right$: ids of left and right neighbors, $\bot$ if undefined \\
\>\>$leaving$: boolean, initially \textbf{false}, read only, application request \\
\>\>$busy$: boolean, initially \textbf{false}; \textbf{true} when servicing a join/leave request or when joining  \\
\>\>$C$: incoming channel\\
\ \\

\>\textbf{actions}\\
\>\emph{joinRequest}:\> \textbf{join} $ \in C \longrightarrow$\\
\>\>\>\textbf{receive} \textbf{join} $(\mathit{reqId})$ \\
\>\>\>$\textbf{if}\ (p < \mathit{reqId} < right)$  \textbf{and}
 \textbf{not} $leaving$ \textbf{and} \textbf{not} $busy$\ \textbf{then}\\
   \>\>\>\>\textbf{send} \textbf{sua}($\mathit{right}$) \textbf{to}  $\mathit{reqId}$ \\
   \>\>\>\>$busy := $ \textbf{true} \\
\>\>\>$\textbf{else} $\\  
\>\>\>\>$\textbf{if}\ \mathit{reqId} < p $ \textbf{then}\\
\>\>\>\>\>\textbf{send} \textbf{join}($\mathit{reqId})$ \textbf{to} $\mathit{left}$ \\
\>\>\>\> $\textbf{else}\ $\\
\>\>\>\>\>\textbf{send} \textbf{join}($\mathit{reqId}$) \textbf{to} $\mathit{right}$ \\

\ \\
\>\emph{leaveRequest}:\> \textbf{leave}$ \ \in C \longrightarrow$\\
\>\>\>\textbf{receive} \textbf{leave}$(\mathit{reqId,q})$ \\
   \>\>\>$\textbf{if}$ $reqId = right$ \textbf{and not} $leaving$ \textbf{and not} $ busy$ \textbf{then}\\
      \>\>\>\>\textbf{send sua}($\mathit{\bot}$) \textbf{to} $\mathit{q}$ \\
      \>\>\>\>$busy :=  \textbf{true} $ \\
\>\>\>$\textbf{else} $\\  
\>\>\>\>$\textbf{if} $ $\mathit{p <= reqId}$ \textbf{then}\\
\>\>\>\>\>\textbf{send} \textbf{leave}($\mathit{reqId}, q)$ \textbf{to} $\mathit{left}$ \\
\>\>\>\> $\textbf{else}\ $\\
\>\>\>\>\>\textbf{send} \textbf{leave}($\mathit{reqId, q}$) \textbf{to} $\mathit{right}$ \\

\ \\
\>\emph{setUpA}:\> \textbf{sua}$ \ \in C \longrightarrow$\\
\>\>\>\textbf{receive} \textbf{sua}$(reqId)$ from $q$   \\
  \>\>\>\textbf{if} $\mathit{reqId \neq \bot}$  \textbf{then} // Join 1.1 received \\
    \>\>\>\> $\mathit{right}$ := $\mathit{reqId}$ \\
      \>\>\>\> $\mathit{left}$ := $\mathit{q}$ \\
    \>\>\>\> \textbf{send sua($\bot$) to} $\mathit{right}$ \\
  \>\>\>\textbf{else} // Join 1.2 or Leave 1 received \\
  \>\>\>\> $\mathit{left}$ := $\mathit{q}$ \\
    \>\>\>\> \textbf{send sub to} $\mathit{left}$ \\

\ \\
\>\emph{setUpB}:\> \textbf{sub}$ \ \in C \longrightarrow$\\
\>\>\>\textbf{receive} \textbf{sub} from $q$   \\
  \>\>\>\textbf{if} $\mathit{q \neq right}$ \textbf{then} // Join 2.2 or Leave 2 received \\
   \>\>\>\>\textbf{send tda} \textbf{to} $\mathit{right}$ \\
   \>\>\>\> $\mathit{right}$ := $q$ \\
   \>\>\>\textbf{else} // Join 2.1 received \\
     \>\>\>\> \textbf{send sub to} $\mathit{left}$ \\
 
\ \\
\>\emph{tearDownA}:\> \textbf{tda}$ \ \in C \longrightarrow$\\
\>\>\>\textbf{receive} \textbf{tda} from $q$  \\
  \>\>\>\textbf{if} $\mathit{q \neq left}$ \textbf{then} // Join 3 or Leave 3.2 received  \\
  \>\>\>\> \textbf{send tdb to} $\mathit{q}$ \\
  \>\>\> \textbf{else} // Leave 3.1 received \\
  \>\>\>\>\textbf{send tda to} $\mathit{right}$ \\
   
\ \\
\>\emph{tearDownB}:\> \textbf{tdb}$ \ \in C \longrightarrow$\\
\>\>\>\textbf{receive} \textbf{tdb} from $q$ \\
  \>\>\>\textbf{if} $\mathit{q \neq right}$ \textbf{then} // Join 4 or Leave 4.2 received \\
  \>\>\>\>\textbf{send ftd to} $\mathit{q}$ \\
  \>\>\>\> $\mathit{busy}$ := \textbf{false} \\
  \>\>\> \textbf{else} // Leave 4.1 received \\
  \>\>\>\>\textbf{send tdb to} $\mathit{left}$ \\
  
  \ \\
  \>\emph{tranDone}:\> \textbf{ftd}$ \ \in C \longrightarrow$\\
  \>\>\>\textbf{receive} \textbf{ftd} from $q$ \\
    \>\>\>\textbf{if} $\mathit{leaving}$ \textbf{then} // Leave 5 received, p may exit \\
    \>\>\>\> $\mathit{right = \bot}$ \\
        \>\>\>\> $\mathit{left = \bot}$ \\
    \>\>\> \textbf{else}  \\
    \>\>\>\>$\mathit{busy :=}$ \textbf{false} // Join 5 received \\

\end{tabbing}
\caption{Algorithm \PROG{UIUC} for process $p$.}\label{figAlgo}
\end{figure}
\clearpage
}

\ \\ \textbf{Algorithm description.} We present a local algorithm
\emph{Unfair Infinite Unlimited Churn} (\PROG{UIUC}) that satisfies
the four properties of the Unfair Infinite Unlimited Churn Problem
while maintaining a linear topology. The basic idea of the algorithm
is to have the \emph{handler} process with the smaller identifier
coordinate churn requests to its immediate right. This handler
considers one such request at a time. This serializes request
processing and guarantees the accepted request's eventual completion.

The algorithm is shown in Figure~\ref{figAlgo}.  To maintain the
topology, each process $p$ has two variables: \emph{left} and
\emph{right} with respective domains less than $p$ and greater than
$p$.  Read-only variable \emph{leaving} is set to \textbf{true} by the
environment once the joined process wishes to leave the overlay
network. Variable \emph{busy} is used by the handler process to
indicate whether it currently coordinates a churn request, or is
initialized to \textbf{true} for a joining process.  The incoming
channel for process $p$ is variable $C$.

The request is sent in the form of a single \emph{join} or
\emph{leave} message.  We assume that a \emph{join} and, for symmetry,
a \emph{leave} message is inserted into an incoming channel of an
arbitrary joined process in the overlay network.

Message \emph{join} carries the identifier of the process wishing to
join the overlay network. Message \emph{leave} carries the id of the
leaving process as well as the id of the process immediately to its
right. Actions \emph{joinRequest} and \emph{leaveRequest} describe the
processing of the two types of requests. If the receiver realizes that
it is to the immediate right of the place of join or leave, and the
receiver is not currently handling another request, i.e. $busy \neq
\textbf{true}$, and it does not want to leave, it starts handling the
arrived request. Otherwise, the recipient process forwards the request
to its left or right.

\begin{figure}[htb]
\centering
\epsfig{figure=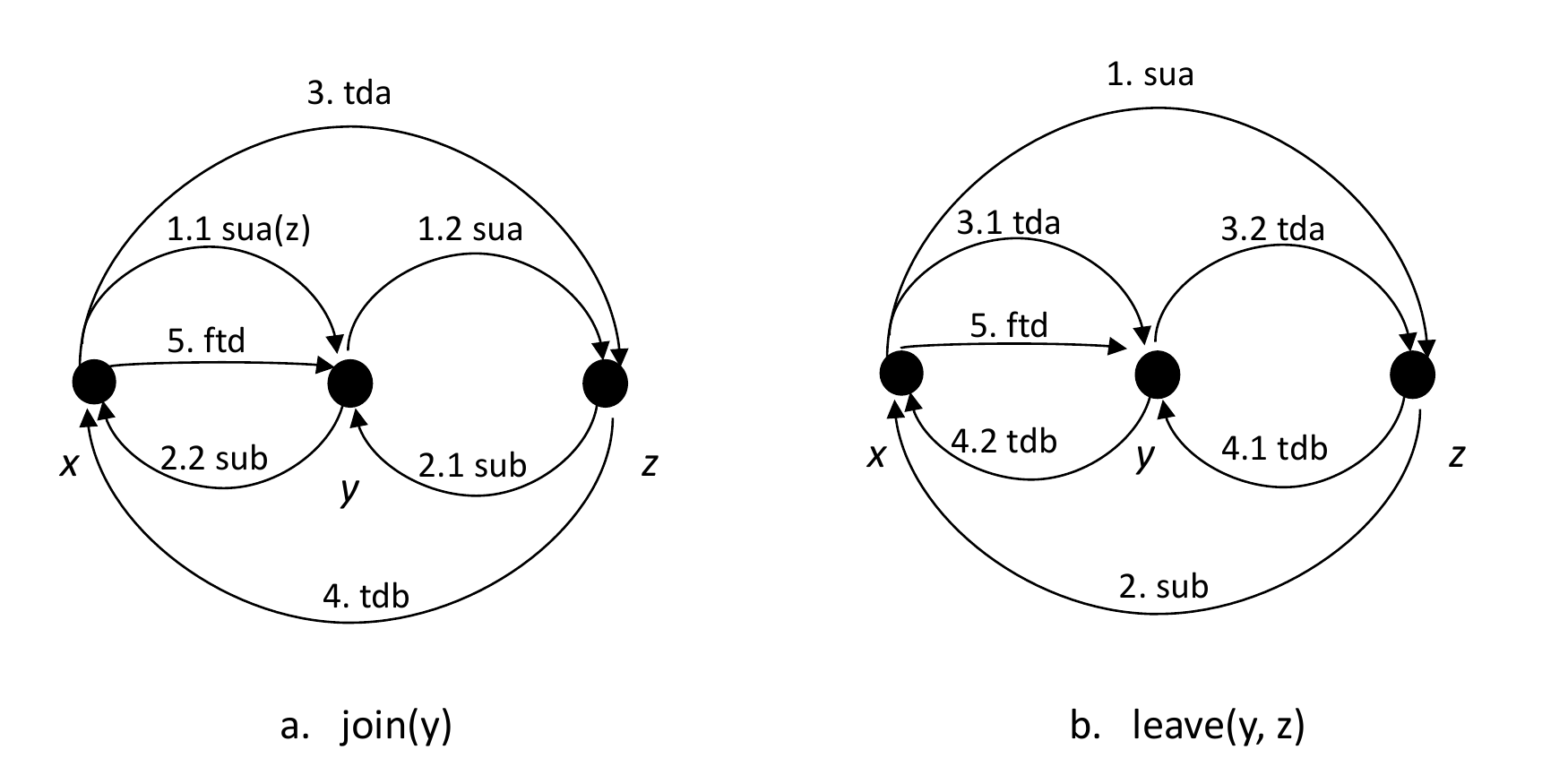, width=\textwidth,clip=}
\caption{\PROG{UIUC} join and leave request handling.}
\label{figJoinLeave}
\end{figure}

Request handling is illustrated in Figure~\ref{figJoinLeave}. It is
similar for join and leave and is divided into five stages.  The first
two stages are \emph{setup} stages: they set up the channels for the
links of the the joining process or for the processes that are the
neighbors of the leaving process. The third and forth stages are
\emph{teardown stages}: they remove the channels of the links being
replaced.  The last stage informs either the leaving process that it
may exit, or the joining process that it may start coordinating its
own churn requests. In the case of join, two links need to be set up,
hence the setup stages are divided into two substages 1.1, 1.2, 2.1
and 2.2. Similarly, in the case of leave, the teardown stages are
divided into substages because two links need to be torn down.  The
messages transmitted during corresponding stages are 1. set up A
\textbf{sua}, 2. set up B \textbf{sub}, 3. tear down A \textbf{tda},
4. tear down B \textbf{tdb} and 5. finish teardown \textbf{ftd}.

\ \\ \textbf{\PROG{UIUC} correctness proof.} We denote message
\textbf{tda} or \textbf{tdb} as \textbf{td*}. Similarly, \textbf{su*}
is \textbf{sua} or \textbf{sub}. 
%
%
Lemmas~\ref{lemMessages} and~\ref{lemStages} follow immediately from
the operation of the algorithm.

\begin{lemma}\label{lemMessages} 
Message \textbf{td*} is the last message in the channel in every
teardown stage. Message \textbf{su*} is the first message in a channel
in every set up stage.
\end{lemma}

\begin{lemma}\label{lemStages} 
The two links of a churning process transition through stages 1
through 5.  No link participates in concurrent transitions.
\end{lemma}

The below corollary follows from Lemma~\ref{lemStages}.

\begin{corollary}\label{lemTerminatingTransition}
Every transitional link is eventually stable.
\end{corollary}

Note that after the link is stable, it may transition again.

\begin{lemma}\label{lemMessageSafety} 
No message in a transitional link is lost.
\end{lemma}
\Proof Observe that according to Lemma~\ref{lemMessages}, in a
teardown stage, the last message in the channel is a \textbf{td*}. By
Lemma~\ref{lemStages}, each stage, including the teardown stages,
eventually completes. By the design of the algorithm, a teardown stage
completes when a \textbf{td*} message is received. Since each channel
is FIFO, this teardown message is received only after all other
messages in the channel are received. That is, after the teardown
stage completes, there are no messages in the channel that is torn
down. According to Lemma~\ref{lemStages}, every channel in a
transitional link is eventually torn down. Hence the lemma.  \QED

\begin{lemma} \label{lemMessageProgress}
Unless the link starts transitioning, a message in a stable link is
either eventually delivered or forwarded to a process closer to
destination.
\end{lemma}

\Proof Consider a message in a channel of a stable link. We assume
fair channel message receipt. This means, unless the channel is
altered due to transitioning, the message is eventually processed
by the recipient process. If this process is the destination, the
message is delivered; if not, the recipient forwards it closer to the
destination.  \QED

\begin{lemma} \label{lemRequestProgress} 
If there is a churn request in the overlay network, some churn request is
eventually satisfied.
\end{lemma}
\Proof Consider state $s_1$ of the computation where there is a churn
request message $m$ in a channel of the overlay network. In this
state, some other churn request processing may be under way. According
to Lemma~\ref{lemTerminatingTransition}, the processing of all these
requests eventually ends. Let $s_2$ be the state of the computation
where all requests under way in $s_1$ are done. If a new request
processing has started in $s_2$, then by applying
Lemma~\ref{lemTerminatingTransition} to it, we obtain the claim of
this lemma.

Let us consider the case where no request processing has occurred
between $s_1$ and $s_2$. According to Lemma~\ref{lemMessageSafety},
message $m$ is sill in some channel of $s_2$. Since no requests are
processed in $s_2$, there are no transitional links. That is, $m$ is
in a stable link. According to Lemma~\ref{lemMessageProgress}, $m$ is
forwarded towards its destination. Continuing this way, we observe
that either $m$ encounters a transitional link or arrives at a process
$p$ that handles this request. If $m$ encounters a transitional link,
this transition started after $s_1$ and the claim of the lemma
follows. If $m$ arrives at $p$, $p$ may either be busy or not. If $p$
is busy, it is handling a request that started after $s_2$. If $p$ is
not busy, it starts processing $m$. Hence the lemma.  \QED

Only processes to the immediate right and left of the churning process
are involved in the processing of the churn request. Hence the
following lemma.

\begin{lemma} \label{lemLocality}
The locality of \PROG{UIUC} is 1.
\end{lemma}


The following theorem follows from
Corollary~\ref{lemTerminatingTransition} and
Lemmas~\ref{lemTerminatingTransition},\ref{lemMessageSafety},\ref{lemMessageProgress},\ref{lemRequestProgress} and~\ref{lemLocality}.

\begin{theorem}\label{theoremUIUC}
Algorithm \PROG{UIUC} is local and it solves the Unfair Infinite Unlimited Churn
Problem.
\end{theorem}

\begin{figure}
\centering
\epsfig{figure=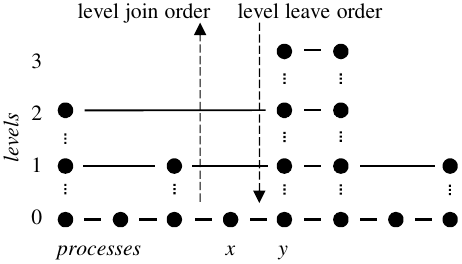,width=12cm,clip=}
\caption{Processes joining and leaving an example skip-list.}
\label{figSkipList}
\end{figure}

\section{\PROG{UIUC} Extensions to Skip List and Skip Graph and Further Work}
Churn algorithm \PROG{UIUC} extends to more complicated topologies
such as skip lists~\cite{corona,pugh1990skip} and skip
graphs~\cite{aspnes2004load,skipgraphsJournal,tiara,goodrich2006rainbow}.
In these topologies, the processes have links on multiple levels.  The
processes are linearized in the lowest level. In the higher levels,
the processes have links to progressively more distant peers. These
higher level links accelerate overlay network searches and other
operations. See Figure~\ref{figSkipList} for an example skip list.  To
extend \PROG{UIUC} to such a structure a separate version of
\PROG{UIUC} should be run at each level. The churn request should bear
the level number to differentiate which level \PROG{UIUC} they belong
to. The churning process should proceed up and down the levels as
follows. A joining process first joins the first, linear, level, then
the next and so on until it joins all the levels appropriate to the
particular structure. The leaving process should proceed in reverse:
the leaving process requests to leave the levels in decreasing
order. For example, in Figure~\ref{figSkipList}, a process $y$ needs
to first leave from level $3$, then $2$ and so on. While a joining
process $x$ needs to first join the overlay network at linearized
level $0$, then proceed to join level $1$ and so on until it reaches
the level appropriate for the particular structure.

As further research it is interesting to consider extensions of
\PROG{UIUC} to ring structures such as Chord~\cite{chord} or
Hyperring~\cite{hyperring}.  Another important area of inquiry is
addition of limited adversarial churn. This problem is difficult to
address in the asynchronous message passing model where the exited
process may not be differentiated from a slow one. Oracles determining
a process exit~\cite{foreback2014stabilizing} may have to be used.

\end{document}